\documentstyle[12pt,a41]{article}

\begin{document}
\thispagestyle{empty}

\mbox{}
\begin{flushleft}
DESY  98-157 \hfill {\tt hep-ph/9810258}\\
September 1998\\
\end{flushleft}
\vspace*{\fill}
\begin{center}
{\LARGE\bf Heavy flavour mass corrections to the}

\vspace{2mm}
{\LARGE\bf longitudinal and transverse cross sections }

\vspace{2mm}
{\LARGE\bf in ${\bf e^+~e^-}$-collisions}

\vspace*{20mm}
\large
{V. Ravindran and W.L. van Neerven \footnote{On leave of absence from 
Instituut-Lorentz, University of Leiden,P.O. Box 9506, 2300 RA Leiden,\\ 
The Netherlands
.}}
\\

\vspace{2em}

\normalsize
{\it DESY-Zeuthen, Platanenallee 6, D-15738 Zeuthen, Germany}%

\vspace*{\fill}
\end{center}
\begin{abstract}
\noindent
We present the heavy flavour mass corrections to the order $\alpha_s$
corrected longitudinal ($\sigma_L$) and transverse ($\sigma_T$)
cross sections in $e^+~e^-$-collisions. Its effect on the value of
the running coupling constant extracted from the longitudinal cross section 
is investigated. Furthermore we make a comparison between the size of
these mass corrections and the magnitude of the order $\alpha_s^2$  
contribution to $\sigma_T$ and $\sigma_L$ which has been recently calculated
for massless quarks. Also studied will be the changes in the above
quantities when the fixed pole mass scheme is replaced by the running 
mass approach.
\end{abstract}

\vspace*{\fill}
Experiments carried out at electron positron colliders like LEP and LSD 
have provided us with a wealth of information about the constants \cite{abc}
appearing in the standard model of the electroweak and strong interactions.
One of them is the strong coupling constant $\alpha_s$ which can be
extracted from various quantities. Examples are the hadronic width of the
Z-boson, the total hadronic cross section, event shapes of jet distributions
and jet rates (for a review on this subject see \cite{elst}). Another
quantity from which $\alpha_s$ can be extracted is the longitudinal
cross section $\sigma_L$ which is measured in the semi inclusive reaction
\begin{eqnarray}
\label{eq1}
e^- + e^+ \rightarrow V \rightarrow {\rm P} + ``X" \,.
\end{eqnarray}
Here V represents the intermediate vector bosons Z and $\gamma$ and $``X"$ 
denotes any inclusive final state. Furthermore P stands for a hadron
which is produced via fragmentation by a quark or a gluon. As we will see
below the perturbation series in QCD for $\sigma_L$
starts in order $\alpha_s$ provided the quark which couples to the
vector boson is massless. Therefore this quantity will become
very sensitive to the value of the strong coupling constant.
Some time ago the perturbation series was only known up to order
$\alpha_s$ (see \cite{alt}, \cite{nawe}). However it turned out that
the result for the longitudinal cross section was much smaller
than its experimental value measured at LEP \cite{busk,akers} which
indicated that the higher order corrections may become rather large.
The latter was confirmed by the second order corrections,
computed very recently in \cite{rijk1}, which amount to about 35 \%
with respect to the lowest order contribution to $\sigma_L$. If
the second order corrections are included the longitudinal cross section
agrees now very well with experiment so that it can be used
to extract $\alpha_s$ (see \cite{abreu1}). Until now
$\sigma_L$ was only evaluated for massless quarks. This will be a correct
approximation for the light quarks like $u,d,s$ and probably also for $c$ but 
it will certainly fail
for the $b$-quark, leave alone for the top quark. Therefore one has to compute
the mass corrections to this quantity at least up to first order in
$\alpha_s$. This contribution will be presented in this paper and we discuss 
its effect on the value of $\alpha_s$ when extracted from experiment via
$\sigma_L$.

The cross section corresponding to process (\ref{eq1}) where the hadron 
P emerges directly, or indirectly via the gluon, from the heavy quark 
H is given by
\begin{eqnarray}
\label{eq2}
  \frac{d^2\sigma^{\rm H,P}(x,Q^2)}{dx\,d\cos\theta} &=& \frac{3}{8}
\left (1+\cos^2\theta\right )\frac{d\sigma_T^{\rm H,P}(x,Q^2)}{dx}
  + \frac{3}{4}\sin^2\theta\frac{d\sigma_L^{\rm H,P}(x,Q^2)}{dx}
\nonumber\\[2ex]
&& + \frac{3}{4}\cos\theta\frac{d\sigma_A^{\rm H,P}(x,Q^2)}{dx} \,.
\end{eqnarray}
~From the equation above one can extract the transverse ($\sigma_T$),
the longitudinal ($\sigma_L$) and the asymmetric ($\sigma_A$) cross sections.
Further $\theta$ denotes the angle between the outgoing hadron P and the
incoming electron. The Bj{\o}rken scaling variable is defined by
\begin{eqnarray}
\label{eq3}
  x = \frac{2pq}{Q^2},\hspace{8mm} q^2 = Q^2 > 0,\hspace{8mm} 0 < x \leq 1\,,
\end{eqnarray}
where the momenta $p$ and $q$ correspond with the outgoing hadron P 
and the virtual vector boson ($Z,\gamma$) respectively.
To obtain the total transverse and longitudinal cross sections
one has to multiply Eq. (\ref{eq2}) with $x$. After integration over $x$
and summation over all hadron species P emerging from the heavy flavour
H one obtains 
\begin{eqnarray}
\label{eq4}
\sigma_k^{\rm H}(Q^2)= \frac{1}{2} \sum_{P} \int_0^1 
dx \, x \, \frac{d~\sigma_k^{\rm H,P}}{dx} (x,Q^2) \qquad k=T,L \,.
\end{eqnarray}
The result above can be written as follows
\begin{eqnarray}
\label{eq5}
\sigma_k^{\rm H}(Q^2) = \sigma_{VV}(Q^2) h_k^v(\rho) + \sigma_{AA}(Q^2)
h_k^a(\rho)  \quad \mbox{with} \quad \rho=\frac{4m^2}{Q^2} \,,
\end{eqnarray}
where $m$ denotes the heavy quark mass. The above definitions for the total
cross sections should not be confused with the ones which are obtained
by integrating $d~\sigma_k/d~x$ without muliplication by $x$. The heavy
flavour contributions to the latter quantities were computed for the first
time in \cite{laer} (see also \cite{abl}-\cite{rane}). The difference
between these two definitions will be pointed out below.
Finally the total cross section for heavy flavour production is given by
\begin{eqnarray}
\label{eq6}
\sigma_{tot}^{\rm H}(Q^2)=\sigma_T^{\rm H}(Q^2)+\sigma_L^{\rm H}(Q^2) \,.
\end{eqnarray}
In Eq. (\ref{eq5}) the pointlike cross sections are
defined by
\begin{eqnarray}
\label{eq7}
  \sigma_{VV}(Q^2) &=& \frac{4\pi\alpha^2}{3Q^2}N\,\left [e_\ell^2 e_q^2 +
\frac{2Q^2(Q^2-M_Z^2)}{\left|Z(Q^2)\right|^2}\,e_\ell e_q C_{V,\ell} C_{V,q} 
\right.
\nonumber\\
&&\left. + \frac{(Q^2)^2}{|Z(Q^2)|^2} 
   \left (C_{V,\ell}^2 + C_{A,\ell}^2\right )C_{V,q}^2 \right ] \,,\\
\label{eq8}
  \sigma_{AA}(Q^2) &=& \frac{4\pi\alpha^2}{3Q^2}N\,\left [
  \frac{(Q^2)^2}{\left|Z(Q^2)\right|^2}
  \left (C_{V,\ell}^2 + C_{A,\ell}^2\right ) C_{A,q}^2 \right ]\,,
\end{eqnarray}
where $N$ denotes the number of colours corresponding to the gauge group
$SU(N)$ (in the case of QCD one has $N=3$).
Furthermore we adopt for the Z-propagator the energy independent
width approximation
\begin{eqnarray}
\label{eq9}
  Z(Q^2) = Q^2 - M_Z^2 + iM_Z\Gamma_Z \,.
\end{eqnarray}
In Eqs. (\ref{eq7}), (\ref{eq8}) the charges of the lepton and the quark 
are denoted by $e_\ell$
and $e_q$ respectively and the electroweak constants are given by
\begin{eqnarray}
\label{eq10}
  \begin{array}{ll}
    C_{A,\ell} = \displaystyle\frac{1}{2\sin2\theta_W}, &
    C_{V,\ell} = -C_{A,\ell}\,(1-4\sin^2\theta_W),\\[2ex]
    C_{A,u} = -C_{A,d} = -C_{A,\ell}, & \\[2ex]
    C_{V,u} = C_{A,\ell}\,\displaystyle(1-\frac{8}{3}\sin^2\theta_W), &
    C_{V,d} = -C_{A,\ell}\,\displaystyle(1-\frac{4}{3}\sin^2\theta_W),
  \end{array}
\end{eqnarray}
The functions $h_k^l$ ($k=T,L$, $l=v,a$) in Eq. (\ref{eq5}) can be
obtained order by order in perturbative QCD from the singlet quark 
and the gluon coefficient functions in the following way 
\begin{eqnarray}
\label{eq11}
h_k^l(\rho)=\int_{\sqrt \rho}^1 dx \,x\,  {\cal C}_{k,q}^{l,\rm S}
(x,\rho,Q^2/\mu^2) + \frac{1}{2} \int_0^{1-\rho} dx \, x \,
{\cal C}_{k,g}^{l,\rm S} (x,\rho,Q^2/\mu^2) \,.
\end{eqnarray}
Here $\mu$ stands for the factorization as well as the renormalization scale.
The result in  Eq. (\ref{eq11}) has been derived from the differential
cross section $d~\sigma_k^{\rm H,P}/d~x$ (see e.g. \cite{nawe}). The latter
can be written as a convolution of the fragmentation densities 
$D_i^{\rm P}$ and the coefficient functions ${\cal C}_{k,i}$ ($i=q,g$)
(see Eq. (2.4) in \cite{rijk2}). Because of the momentum sum rule
satisfied by the fragmentation densities (see Eq. (2.9) in \cite{nawe})
expression (\ref{eq11}) follows immediately. Notice that the functions
$h_k^l$ differ from the functions $f_k^l$ computed
in \cite{rane} (see also the functions $H_i$ ($i=2,6$) in \cite{abl}).
The latter were derived from the first moment of the
non-singlet quark coefficient function and they differ from the former
which are given by the second moment of
the singlet quark and gluon coefficient functions presented in Eq. (\ref{eq11}).
However because of Eq. (\ref{eq5}) the sum of the transverse and longitudinal
components of both functions have to lead to the same 
perturbation series of the total cross section so that one has the relation 
$h_T^l+h_L^l=f_T^l+f_L^l$ ($l=v,a$).
The functions $h_k^l$ can be expanded in the strong coupling constant
$\alpha_s$ as follows
\begin{eqnarray}
\label{eq12}
h_k^l(\rho) = \sum_{n=0}^{\infty} \left (\frac{\alpha_s(\mu)}{4\pi} \right )^n
h_k^{l,(n)}(\rho) \,.
\end{eqnarray}
The lowest order contributions corresponding to the Born reaction 
\begin{eqnarray}
\label{eq13}
V \rightarrow {\rm H} + {\overline {\rm H}} \,,
\end{eqnarray}
with $V= \gamma, Z$ are given by
\begin{eqnarray}
\label{eq14}
h_T^{v,(0)}(\rho) &=& \sqrt{1- \rho} \qquad h_L^{v,(0)}(\rho) \,,
=\frac{\rho}{2} \sqrt{1- \rho}
\nonumber\\[2ex]
h_T^{a,(0)}(\rho) &=& (1- \rho)^{3/2} \qquad h_L^{a,(0)}(\rho)=0 \,,
\end{eqnarray}
where $\rho$ is defined in Eq. (\ref{eq5}). The next-to-leading order (NLO)
contributions originate from the one-loop virtual corrections to reaction
(\ref{eq13}) and the gluon bremsstrahlungs process
\begin{eqnarray}
\label{eq15}
V \rightarrow {\rm H} + {\overline {\rm H}} + g\,.
\end{eqnarray}
We have computed the order $\alpha_s$ contributions to the quark
${\cal C}_{k,q}^{l,\rm S}$ and gluon ${\cal C}_{k,g}^{l,\rm S}$
coefficients functions for the case $m\not=0$ in Eq. (\ref{eq11}) 
and confirmed the result obtained in appendix A of \cite{nawe}. Notice
that the quark coefficient functions are derived from the process where
the hadron P emerges from the quark H. The gluon coefficient function
corresponds with the process where the hadron is emitted by the gluon.
After performing the integral in  Eq. (\ref{eq11}) we obtain the following 
results for $k=T,L$
\begin{eqnarray}
\label{eq16}
h_T^{v,(1)}(\rho)&=&C_F \left [\frac{1}{2}\rho(1 - 3 \rho)F_1(t)+ \rho^{3/2}
(1 + \rho)
F_2(t)+(32 - \frac{39}{2}\rho + \frac{21}{2} \rho^2) {\rm Li}_2(t)
\right.
\nonumber\\
&& \left. +(16 - 10 \rho + 6\rho^2)F_3(t)+2 \sqrt {1-\rho}F_4(t)
+(8 - 6\rho + 6\rho^2)\ln(t) \ln(1+t) \right.
\nonumber\\
&& \left. +(-12 + 3 \rho - \frac{3}{2} \rho^2)\ln(t)
- 5 \rho \sqrt {1-\rho} \right ] \,,\\
\label{eq17}
h_L^{v,(1)}(\rho)&=&C_F \left [-\frac{1}{2}\rho(1-3 \rho)F_1(t)
- \rho^{3/2}(1+\rho) F_2(t)+(\frac{39}{2}\rho - \frac{37}{2} \rho^2) 
{\rm Li}_2(t) \right.
\nonumber\\
&& \left. + 10 \rho (1 - \rho)F_3(t)+ \rho \sqrt {1-\rho}F_4(t)
+( 6\rho- 8\rho^2) \ln(t) \ln(1+t) \right.
\nonumber\\
&& \left. +(- \rho + \frac{13}{4} \rho^2)\ln(t)
+  (3 + \frac{19}{2} \rho)\sqrt {1-\rho} \right ]\,,\\
\label{eq18}
 h_T^{a,(1)}(\rho)&=&C_F \left [\frac{1}{2}\rho(1 - 4 \rho)F_1(t)
+ \rho^{3/2}(1 + 2\rho)
F_2(t)+(32 - \frac{103}{2}\rho + 30 \rho^2) {\rm Li}_2(t) \right.
\nonumber\\
&& \left. +(16 - 26 \rho + 16\rho^2)F_3(t)+2 (1 - \rho)^{3/2}F_4(t)
+(8 - 14\rho + 12 \rho^2)\ln(t) \ln(1+t) \right.
\nonumber\\
&& \left. +(-12 + 9 \rho - \frac{9}{2} \rho^2+\frac{3}{4}\rho^3 )\ln(t)
- (12 \rho + \frac{3}{2} \rho^2) \sqrt {1-\rho} \right ]\,,\\
\label{eq19}
h_L^{a,(1)}(\rho)&=&C_F \left [-\frac{1}{2}\rho(1 - 4 \rho)(F_1(t)- 4 F_3(t)
 - 7{\rm Li}_2(t) - 4 \ln(t) \ln(1+t) )- \rho^{3/2}(1 + 2\rho) F_2(t) \right.
\nonumber\\
&& \left. +(2 \rho +  \frac{13}{4} \rho^2 - \frac{9}{8} \rho^3)\ln(t)
+  (3 + 3 \rho + \frac{9}{4} \rho^2)\sqrt {1-\rho} \right ]\,,
\end{eqnarray}
with
\begin{eqnarray}
\label{eq20}
t=\frac{1-\sqrt{1-\rho}}{1+\sqrt{1-\rho}}
\end{eqnarray}
and the colour factor $C_F$ is given by $C_F=(N^2-1)/2N$. 
The functions $F_i(t)$ appearing above are defined by
\begin{eqnarray}
\label{eq21}
 F_1(t)&=&{\rm Li}_2(t^3)+4\zeta(2)+\frac{1}{2}\ln^2(t)+3\ln(t)\ln(1+t+t^2)
\,\\
\label{eq22}
 F_2(t)&=&{\rm Li}_2(-t^{3/2})-{\rm Li}_2(t^{3/2})+{\rm Li}_2(-t^{1/2})
-{\rm Li}_2(t^{1/2}) + 3 \zeta(2) + 2\ln(t)\ln(1 + \sqrt t)
\nonumber\\
&& -2\ln(t)\ln(1 - \sqrt t) + \frac{3}{2}\ln(t)\ln(1+t-\sqrt t)
 -\frac{3}{2}\ln(t)\ln(1+t+\sqrt t) \,\\
\label{eq23}
 F_3(t)&=&{\rm Li}_2(-t) + \ln(t)\ln(1 - t) \,\\
\label{eq24}
 F_4(t)&=&6 \ln(t) - 8\ln(1 - t) - 4\ln(1 + t) \,,
\end{eqnarray}
where $\zeta(n)$, which occurs in the formulae of this paper for $n=2,3$,
represents the Riemann $\zeta$-function and ${\rm Li}_2(x)$ denotes the
dilogarithm.
Using Eqs. (\ref{eq16}) - (\ref{eq19}) one can check that the order $\alpha_s$
contribution to the total cross section in (\ref{eq6}) is in agreement with
the literature \cite{stn}.  
The next-to-next-to-leading
order (NNLO) contributions come from the following processes. First one has to
calculate the two-loop vertex corrections to the Born process (\ref{eq13}) and
the one-loop corrections to (\ref{eq15}). Second one has to add the radiative
corrections due to the following reactions
\begin{eqnarray}
\label{eq25}
V \rightarrow {\rm H} + {\overline {\rm H}} + g + g
\end{eqnarray}
\begin{eqnarray}
\label{eq26}
V \rightarrow {\rm H} + {\overline {\rm H}} + {\rm H} + {\overline {\rm H}}
\end{eqnarray}
\begin{eqnarray}
\label{eq27}
V \rightarrow {\rm H} + {\overline {\rm H}} + q + {\overline q}
\end{eqnarray}
where $q$ denotes the light quark. The results for $h_k^{l,(2)}$ presented 
below are only computed for those contributions containing Feynman graphs where
the vector boson V is always coupled to the heavy quark H so that the 
light quarks can be
only produced via fermion pair production emerging from a gluon splitting.
Because we are only interested in the ratios 
\begin{eqnarray}
\label{eq28}
R_k(Q^2) = \frac{\sigma_k^{\rm H}(Q^2)}{\sigma_{tot}^{\rm H}(Q^2)}
\quad \mbox{for} \quad k=T,L
\end{eqnarray}
we do not consider
other contributions which drop out in the expression above provided we put the
mass of H equal to zero in the second order correction.
The contributions which can be omitted are given by
one- and two-loop vertex corrections
which contain the triangular quark-loop graphs \cite{knku}, \cite{ghp}. 
They only show up if the quarks
are massive and are coupled to the Z-boson via the axial-vector vertex. Notice
that one has to sum over all members of one quark family in order to cancel
the anomaly. 
Further we exclude all contributions from reaction (\ref{eq27}) which involve 
Feynman graphs where the light quarks q are coupled to the vector boson V. 
Notice that interference terms of the latter with diagrams where heavy quark 
are attached to the vector boson vanish if the heavy quark is taken to be
massless provided one sums over all members in one family.

The order $\alpha_s^2$ contributions to the quark and gluon coefficient
functions have been calculated in \cite{rijk1}, \cite{rijk2}. Because of the
complexity of the calculation of these functions the heavy quark mass was taken 
to be zero.
This approximation is good for the charm and bottom quark but not for the top
quark as we will see below. Substituting these coefficient functions in the
integrand of Eq. (\ref{eq11}) (here $\rho=0$) we obtain
\begin{eqnarray}
\label{eq29}
h_T^{v,(2)}=h_T^{a,(2)}&=&C_F^2 \left \{ 6 \right \} +C_AC_F \left \{
 - \frac{89}{15} - \frac{196}{5} \zeta(3) \right \}
  +n_f C_F T_f \left \{  \frac{8}{3} + 16 \zeta(3) \right \} \,\\
\label{eq30}
 h_L^{v,(2)}=h_L^{a,(2)}&=&C_F^2 \left \{ -\frac{15}{2} \right \} 
+C_AC_F \left \{- 11 \ln \left ( \frac{Q^2}{\mu^2}\right )
+\frac{2023}{30}-\frac{24}{5} \zeta(3)\right \}
\nonumber\\
&& +n_f C_F T_f \left \{4 \ln \left ( \frac{Q^2}{\mu^2}\right ) 
- \frac{74}{3} \right \}
\end{eqnarray}
Where the colour factors are given by $C_A=N$ and $T_f=1/2$ (for $C_F$
see below Eq. (\ref{eq20})). Further $n_f$ 
denotes the number of light flavours which originate from process (\ref{eq27}).
Finaly $\mu$ appearing in the strong coupling constant $\alpha_s$ and
the logarithms in Eq. (\ref{eq30}) represents the
renormalization scale. Notice that the coefficient of the logarithm is
proportional to the lowest order coefficient of the $\beta$-function.
The logarithm does not appear in $h_T^{l,(2)}$
because $h_T^{l,(1)}=0$ for $m=0$ and $l=v,a$.
Since $m=0$ there is no distinction between $h_k^{v,(2)}$ and 
$h_k^{a,(2)}$
for $k=T,L$ anymore unlike in the case of the first order corrections
in Eqs. (\ref{eq16})-(\ref{eq19}) where the heavy quark was taken to be massive.
Furthermore one can check that substitution of Eqs. (\ref{eq29}), (\ref{eq30})
into Eq. (\ref{eq6}) provides us with the order $\alpha_s^2$ contribution
to the total cross section which is in agreement with the  
results obtained in \cite{ckt}. 
When the quark is massless we get from Eqs. (\ref{eq5}), (\ref{eq6})
the following perturbation series for the ratio
in Eq. (\ref{eq28}) up to order $\alpha_s^2$
\begin{eqnarray}
\label{eq31}
R_L^{\rm H}(Q^2)&=&  
\frac{\alpha_s(\mu)}{4\pi} C_F \left \{ 3 \right \} +
\left (\frac{\alpha_s(\mu)}{4\pi}\right )^2 \left [ C_F^2 \left \{-\frac{33}{2} 
\right \}+ C_AC_F \left \{- 11 \ln\left (\frac{Q^2}{\mu^2}\right ) 
+ \frac{123}{2} - 44 \zeta(3) \right \} \right.
\nonumber\\[2ex]
&& \left. + n_fC_F T_F \left \{4 \ln\left(\frac{Q^2}{\mu^2}\right ) 
- \frac{74}{3} \right \} \right ] \,,\\
\label{eq32}
R_T^{\rm H}(Q^2)&=& 1 - R_L^{\rm H}(Q^2)
\end{eqnarray}
We will now show how the results for the above ratios will be
modified when the Born and the first order contribution are
computed for massive quarks. Further we discuss the consequences 
of our findings for the extraction of $\alpha_s$.
In order to make the comparison between the massless and massive approach
to Eq. (\ref{eq28}) we have chosen the following
parameters (see \cite{caso}). The electroweak constants are: $M_Z = 91.187
~{\rm GeV/c^2}$, $\Gamma_Z = 2.490~{\rm GeV/c^2}$ and $\sin^2 \theta_W =
0.23116$. For the strong parameters we choose : $\Lambda_5^{{\overline {MS}}}
=237~{\rm MeV/c}$ ($n_f=5$) which implies $\alpha_s(M_Z)=0.119$ (two-loop
corrected running coupling constant). Further we take for the renormalization
scale $\mu=Q$ unless mentioned otherwise. Notice that we study
$R_k^{\rm H}$ for ${\rm H}=c,b$ at the CM energy $Q=M_Z$. For the
heavy flavour masses the following values are adopted : $m_c= 1.5
~{\rm GeV/c^2}$,  $m_b=~4.50~{\rm GeV/c^2}$ and $m_t= 173.8~{\rm GeV/c^2}$. The
results for the bottom quark can be found in table \ref{fig1}.
A comparison between the first and second column reveals that on the
Born level the diference between the massive and massless approach to
$R_T^b$ is very small (about two promille). For $R_L^b$ it is more conspicious
but the correction due to mass effects, which equals 0.0014 for $R_L^{(1)}$,
is still much smaller than the order $\alpha_s^2$ contribution which amounts
to 0.0121. The corrections due to mass effects are also smaller than the
changes caused by a different choice of the renormalization scale. If we choose
$\mu=Q/2$ or $\mu=2~Q$ one gets $R_L^{(2)}=0.0509$ and $R_L^{(2)}=0.0461$
respectively which differ by 0.0024 from the central value $R_L^{(2)}=0.0485$.
\begin{table}
\begin{center}
\begin{tabular}{|c|c|c|c|}\hline
\multicolumn{4}{|c|}{$R_T^b$} \\ \hline \hline
& $m_b=0.0~{\rm GeV/c^2}$& $m_b=4.50~{\rm GeV/c^2}$&
$\overline{m}_b(M_Z)=2.80~{\rm GeV/c^2}$ \\ \hline \hline
$R_T^{(0)}$ & 1.0         & 0.999      & 0.999            \\
$R_T^{(1)}$ & 0.962       & 0.964      & 0.963  \\
$R_T^{(2)}$ & 0.950       & 0.952      & 0.951  \\ \hline \hline
\multicolumn{4}{|c|}{$R_L^b$} \\ \hline \hline
$R_L^{(0)}$ & 0.0         &  0.0016     &  0.0006  \\ 
$R_L^{(1)}$ & 0.0378      &  0.0364     &  0.0374  \\
$R_L^{(2)}$ & 0.0499      &  0.0485     &  0.0495  \\ \hline
\end{tabular}
\end{center}
\caption{The ratio $R_k=\sigma_k/\sigma_{tot}$ ($k=T,L$) for 
$b \bar b$-production.}
\label{fig1}
\end{table}
We also studied the effect of the running quark mass on the ratio in
Eq. (\ref{eq28}). For this purpose one has to change
the on-mass shell scheme used in Eqs. (\ref{eq16})-(\ref{eq19})
into the ${\overline {\rm MS}}$-scheme. This can be done by substituting in
all expressions the fixed pole mass $m$ by the running mass ${\overline {m}}
(\mu)$. Moreover one has to add to the first order contributions
(\ref{eq16})-(\ref{eq19}) the finite counter term
\begin{eqnarray}
\label{eq33}
\Delta h_k^{l,(1)} = {\overline {m}}(\mu) C_F \left [ 4 -
3 \ln \left (\frac{{\overline {m}}^2(\mu)}{\mu^2}\right ) \right ]
\left (\frac{d~h_k^{l,(0)}(\rho)}{d~m} \right )_{m={\overline {m}}(\mu)} \,,
\end{eqnarray}
where $\mu$ stands for the mass renormalization scale for which we choose
$\mu=Q$.
Further we adopt the two-loop corrected running mass with the initial
condition $\overline {m}(\mu_0)=\mu_0$. Using the relation between
the $\overline{MS}$-mass and the fixed pole mass, as is indicated by the first
factor on the righthand side in Eq. (\ref{eq33}), we have taken for bottom
production $\mu_0=4.10~{\rm GeV/c^2}$ which corresponds with a pole mass
$m_b=4.50~{\rm GeV/c^2}$. This choice leads to
${\overline {m}}_b(M_Z)=2.80~{\rm GeV/c^2}$ 
which is 5 \% above the experimental value 2.67 ${\rm GeV/c^2}$ measured at 
LEP \cite{abreu2}.  The results are presented in the third column
of table \ref{fig1}. Comparing the second with the third column we observe
that the mass corrections decrease because the running mass
is smaller than the fixed pole mass. It is now interesting to see how
the mass terms contributing to $R_L^b$ affect the extraction of
the running coupling constant at $\mu=Q=M_Z$. To that order we equate
the mass corrected formula for $R_L^b$ with the massless result presented 
in Eq. (\ref{eq31}). The latter yields $R_L^{(2)}=0.0499$ for
$\alpha_s=0.119$ (see first column, third row of table \ref{fig1}). 
Using the same number for the massive expression of $R_L^{(2)}$ we obtain 
$\alpha_s=0.122$ which amounts to an enhancement of 2.5 \%  for 
$m_b=4.50~{\rm GeV/c^2}$.
In the case of a running mass i.e. ${\overline {m}}_b(M_Z)=2.80~{\rm GeV/c^2}$ 
we get $\alpha_s=0.120$ so that the enhancement becomes 1.0 \%.
However as we have said before these mass effects are
smaller than those caused by a variation in the renormalization scale.
Following the same procedure we equate $R_L$ at different values of
$\mu$. Choosing $\mu=Q$ one infers from table \ref{fig1} that 
$R_L^{(2)}=0.0499$ for $\alpha_s=0.119$. This value of $R_L^{(2)}$ can also
be obtained at $\mu=Q/2$ but then one has to take $\alpha_s=0.127$.
If we choose $\mu=2~Q$ the result is
$\alpha_s=0.112$. Therefore the uncertainty is about 6\% which is larger
than the mass correction. The latter becomes even smaller when we also add
the part coming from the bottom quark to the longitudinal cross section 
consisting of the contributions of the
light quarks and the charm quark. Concerning the latter
quark we want to remark that its mass is so small that the massive
approach will become indistinguishable from the massless results.
Since the mass effects up to the order $\alpha_s$ level are rather small
even for the bottom quark we can assume that the second order corrections,
derived for $m=0$, also apply for $m\not=0$ at least as long as
$Q\gg m$. This will be correct at large collider energies for all 
quarks except for the top quark as we will see below.

For $t \bar t$ production we will study the mass effects up to order 
$\alpha_s$ at a CM energy $Q=500~{\rm GeV/c}$. Here we have chosen
$\mu_0=166.1~{\rm GeV/c^2}$ which corresponds with a 
fixed pole mass of $m_t=173.8~{\rm GeV/c^2}$.  This choice leads to
${\overline {m}}_t(Q)=153.5~{\rm GeV/c^2}$.
From table \ref{fig2} we infer that for this process the mass corrections are 
huge and they are much larger than 
the first and second order corrections. Therefore the numbers in the last row,
presented for $R_T^t$ and $R_L^t$, are unreliable because they only
hold if the mass corrections in the order $\alpha_s^2$ contributions can
be neglected as was done in Eqs. (\ref{eq29}), (\ref{eq30}). This implies
that for the top quark $R_T^{(2)}$ and $R_L^{(2)}$ have to be computed for 
$m\not =0$ which will be an enormous
enterprise. We also studied the effect of the running mass presented
in the third column of table \ref{fig2}. Here the difference 
between the fixed pole mass and the running mass approach is
larger than the one observed for the bottom quark in table \ref{fig1}. Another
feature is that the order $\alpha_s$ correction increases when the
running mass scheme is used and its sign is reversed with respect
to the correction obtained for the fixed pole mass approach. Notice that a
study of the change in $R_k^t$ ($k=T,L$) under variation of the renormalization
scale makes no sense because of the missing exact result in second order for
$m\not =0$.
\begin{table}
\begin{center}
\begin{tabular}{|c|c|c|c|}\hline
\multicolumn{4}{|c|}{$R_T^t$} \\ \hline \hline
&$m_t=0~{\rm GeV/c^2}$&$m_t=173.8~{\rm GeV/c^2}$&
$\overline{m}_t(Q)=153.5~{\rm GeV/c^2}$ \\ \hline \hline
$R_T^{(0)}$ & 1.0         & 0.826      & 0.862  \\
$R_T^{(1)}$ & 0.970       & 0.828      & 0.834  \\ 
$R_T^{(2)}$ & 0.962       & 0.821      & 0.826  \\ \hline \hline  
\multicolumn{4}{|c|}{$R_L^t$} \\ \hline \hline
$R_L^{(0)}$ & 0.0         &  0.175     &  0.138  \\
$R_L^{(1)}$ & 0.030       &  0.172     &  0.167  \\ 
$R_L^{(2)}$ & 0.038       &  0.180     &  0.174   \\ \hline
\end{tabular}
\end{center}
\caption{The ratio $R_k=\sigma_k/\sigma_{tot}$ ($k=T,L$) for
$t \bar t$-production at $Q=500~{\rm GeV/c}$.}
\label{fig2}
\end{table}

Summarizing our findings we have computed the mass corrections up
order $\alpha_s$ for the transverse and longitudinal
cross sections which are due to heavy flavours. In the case of charm
there is no observable difference between the massless 
and massive approach. For the bottom quark we
see a difference which leads to an enhancement of the extracted
value of $\alpha_s$ which is of the order of 2.5 \% (fixed pole mass 
$m_b=4.50~{\rm GeV/c^2}$).
or 1.0 \%  (running mass $\overline{m}_b(M_Z)=2.80~{\rm GeV/c^2}$).
These numbers become much smaller
when the light flavour contributions are added
to the cross sections. A variation in the renormalization scale
introduces larger effects on the value for $\alpha_s$ which are 
about 6.0 \% (see e.g also \cite{abreu1}). In the case of top-quark production
the mass terms cannot be neglected anymore since they are much larger
than the second order corrections computed for massless quarks.\\[5mm]
\noindent

Acknowledgements\\

The authors would like to thank J. Bl\"umlein for reading the manuscript
and giving us some useful remarks.
This work is supported by the EC-network under contract FMRX-CT98-0194.

\end{document}